# Chapter 6

# Semi-metric Behavior in Document Networks and its Application to Recommendation Systems

## Luis M. Rocha

Recommendation systems for different Document Networks (DN) such as the World Wide Web (WWW), Digital Libraries, or Scientific Databases, often make use of distance functions extracted from relationships among documents and between documents and semantic tags. For instance, documents in the WWW are related via a hyperlink network, while documents in bibliographic databases are related by citation and collaboration networks. Furthermore, documents can be related to semantic tags such as keywords used to describe their content. The distance functions computed from these relations establish associative networks among items of the DN, referred to as Distance Graphs, which allow recommendation systems to identify relevant associations for individual users. However, modern recommendation systems need to integrate associative data (defined by distance graphs) generated from multiple sources such as different databases, web sites, and even other users. Thus, we are presented with a problem of combining evidence (about associations between items) from different sources characterized by distance functions. In this paper we describe our work on (1) inferring relevant associations from, as well as characterizing, semi-metric distance graphs and (2) combining evidence from different distance graphs in a recommendation system. Regarding (1), we present the idea of semi-metric distance graphs, and introduce ratios to measure semi-metric behavior. We compute these ratios for several DN such as digital libraries and web sites and show that they are useful to identify implicit associations. We also propose a model based on semi-metric behavior that allow us to quantify the amount of important latent associations in a DN. Regarding (2), we describe an algorithm to combine evidence from distance graphs that uses Evidence Sets, a set structure based on Interval Valued Fuzzy Sets and Dempster-Shafer Theory of Evidence. This algorithm has been developed for a recommendation system named *TalkMine*.

# 1. Document Networks and Recommendation Systems

The prime example of a Document Network (DN) is the World Wide Web (WWW). But many other types of such networks exist: bibliographic databases containing scientific publications, preprints, internal reports, as well as databases of datasets used in scientific endeavors[1]. Each of these databases possesses several distinct relationships among documents and between documents and semantic tags or indices that classify documents appropriately.

DN typically function as information resources for communities of users who query them to obtain relevant information for their activities. We often refer to collections of document networks and communities of users as *Distributed Information Systems* (DIS) (Rocha 2001b). DIS such as the Internet, Digital Libraries, and the like have become ubiquitous in the past decade, demanding the development of new techniques to both cater to the information needs of communities of users as well as to understand several aspects of the structure and dynamics of DN. The first set of techniques, needed to respond to engineering needs, come from the field of Information Retrieval, and are typically known as Recommender Systems (Krulwich & Burkey 1996, Konstan et al. 1997, Herlocker et al. 1999, Rocha 2001b). The second set of techniques, needed to respond to scientific interest in DN, come from the desire to analyze networks of documents and/or communities and is more related to Graph theory, Algebra, Complex Systems, as well as Linguistics (Kleinberg 1999, Chakrabarti et al. 1999, Berry, Dumais, & Obrien 1995, Li, Burgess, & Lund 2000, Newman 2001). Clearly, the scientific and engineering techniques complement and influence each other. We expect to use scientific knowledge about DN to improve recommendation algorithms, namely by better understanding users and information resources, and by producing adaptive DN, or adaptive webs (Bollen & Heylighen 1998, Bollen & Rocha 2000, Rocha 2001a).

---

[1]Such as MEDLINE (http://www.nlm.nih.gov), the e-Print Arxiv (http://xxx.lanl.gov/), and the GenBank (http://www.ncbi.nlm.nih.gov/Genbank/) for Nucleic Acid Sequences.

The information retrieval and recommendation systems we have developed in this area are based on Multi-Agent algorithms which integrate knowledge about the association amongst elements of DN, amongst users, and about the interests of individual users and their communities. In particular, a soft computing algorithm (*TalkMine*) has been created to integrate such evidence and in so doing adapt DN to the expectations of their users (Rocha 2001b). The process of integration of knowledge in *TalkMine* requires the construction of distance graphs from DN that characterize the associations amongst their components. We summarize in sections 8 and 9 how *TalkMine* uses evidence theory and fuzzy set constructs to integrate such distance graphs to obtain measurements of relevance and produce recommendations. Before that, we detail the main goal of the present work, which is to study distance graphs extracted from DN. We show that their metric behavior can be used (1) as an indicator of the relevance of collections of documents and the interests of users who have selected certain sets of documents, (2) to identify the trends in communities associated with sets of documents, and (3) to study the characteristics of such DN in general, particularly, the amount of important latent associations.

Our approach to the third of these goals, in particular, is based on empirical evidence accumulated from several real and artificial DN. It is also independent of the particularities of *TalkMine* or any other specific recommendation algorithm. Indeed, our approach aims at a general characterization of how associative knowledge is stored in DN. Goals 1 and 2 are detailed in sections 5 and 6. Goal 3 is detailed in section 7. Sections 2 to 4 outline the foundations of semi-metric behavior required for later sections

# 2. Characterizing Document Networks with Distance Functions

## 2.1 Harvesting Relations from Document Networks

For each DN we can identify several distinct relations among documents and between documents and semantic tags used to classify documents appropriately. For instance, documents in the WWW are related via a hyperlink network, while documents in bibliographic databases are related by citation and collaboration networks (Newman 2001). Furthermore, documents are typically related to semantic tags such as keywords used to describe their content. Although all the technology here discussed would apply equally to any of these relations extracted from DN, let us exemplify the problem with the datasets we have used in the *Active Recommendation Project* (ARP) (http://arp.lanl.gov), part of the Library Without Walls Project, at the Research Library of the Los Alamos National Laboratory (Bollen & Rocha 2000).

ARP is engaged in research and development of recommendation systems for digital libraries. The *information resources* available to ARP are large databases with academic articles. These databases contain bibliographic, citation, and sometimes abstract information about academic articles. One of the databases we work with is *SciSearch*, containing articles from scientific journals from several fields collected by ISI (Institute for Scientific Indexing). We collected all SciSearch data from the years of 1996 to 1999. There are 2,915,258 records[2], from which we extracted 839,297 keywords (semantic tags) that occurred at least in two distinct documents. The sources of keywords are the terms authors and/or editors chose to categorize (index) documents, as well as title words. Note that these <u>do not</u> include any other words that may occur in the text of the

---

[2]Records contain bibliographical information about published documents. Records can be thought of as unique pointers to documents, thus, for the purposes of this article, the two terms are interchangeable.

articles – only words specified as keywords and title words were included in this study. We removed typical stop words and stemmed all remaining keywords. The 10 most common (stemmed) keywords in the ARP dataset are listed in Table 1.

The relation between documents and keywords allows us to infer the semantic value of documents and the inter-associations between keywords. Naturally, semantics is ultimately only expressed in the brains of users who utilize the documents, but keywords are symbolic tokens of this ultimate expression, which we can try to infer from the relation between documents and keywords. Such semantic relation is stored as a very sparse *Keyword-Document Matrix* **A**. Each entry $a_{i,j}$ in the matrix is boolean and indicates whether keyword $k_i$ indexes ($a_{i,j} = 1$) document $d_j$ or not ($a_{i,j} = 0$).

Table 1: 10 Most Common (stemmed) Keywords and their frequency

| Frequency | Keyword |
|-----------|---------|
| 188498    | cell    |
| 144294    | system  |
| 140258    | studi   |
| 138128    | express |
| 129679    | protein |
| 122587    | model   |
| 116900    | activ   |
| 112538    | rat     |
| 107240    | gene    |
| 106497    | human   |

The structure of a DN is likewise defined by the relations between documents in the document collection. In academic databases these relations refer to citations, while in the WWW to hyperlinks. In subsequent sections we work mostly with the semantic relation of DN (as defined here), but in section 7 we also work with the hyperlink structure of web sites.

## 2.2 Computing Distance Functions: Associative Semantics

To discern closeness between keywords according to the documents they classify, we compute the *Keyword Semantic Proximity* (*KSP*), obtained from **A** by the following formula:

$$KSP(k_i, k_j) = \frac{\sum_{k=1}^{m}(a_{i,k} \wedge a_{j,k})}{\sum_{k=1}^{m}(a_{i,k} \vee a_{j,k})} = \frac{N_\cap(k_i, k_j)}{N(k_i) + N(k_j) - N_\cap(k_i, k_j)} \quad (1)$$

The semantic proximity[3] between two keywords, $k_i$ and $k_j$, is the probability that keywords $k_i$ and $k_j$ co-index the same document in a DN whose semantics is defined by matrix **A**. It depends on the sets of documents indexed by each keyword, and the intersection of these sets. $N(k_i)$ is the number of documents keyword $k_i$ indexes, and $N_\cap(k_i, k_j)$ the number of documents both keywords index. This last quantity is the number of elements in the intersection of the sets of documents that each keyword indexes. Thus, two keywords are near if they tend to index many of the same documents.

From the inverse of *KSP* we obtain a distance function between keywords:

$$d(k_i, k_j) = \frac{1}{KSP(k_i, k_j)} - 1 \quad (2)$$

*d* is a distance function because it is a nonnegative, symmetric, real-valued function such that $d(k, k) = 0$ (Shore & Sawyer 1993). It defines a weighted graph *D*, which we refer to as a *distance graph*, whose vertices are all of the keywords extracted from a given DN, and the edges are the values of $d(k_i, k_j)$ for pairs of keywords $(k_i, k_j)$. A small distance between keywords implies a strong semantic association between keywords, in the case of the ARP dataset, inferred from the probability of co-indexing documents. This way, this distance function indicates how far, semantically, a keyword is from another given a specific set of documents. In this sense, the *associative semantics* captured by *d* is context-dependent, as discussed next.

---

[3]This measure of closeness, formally, is a proximity relation (Klir & Yuan 1995, Miyamoto 1990) because it is a reflexive and symmetric fuzzy relation. Its transitive closure is known as a similarity relation (Ibid).

## 2.3 Characterizing Information Resources and Users

Clearly, many other types of distance functions can be defined on the elements of a DN. Naturally, the conclusions drawn cannot be separated by how well, and how appropriately for a given application, a distance function is capable of discerning the elements of the set it is applied to. Thus, distance functions applied to citation structures or collaboration networks would require distinct semantic considerations than those used for keyword sets.

In any case, we characterize an information resource with sets of distance functions such as formula 2. We assume that the collection of all relevant associative distance graphs extracted from a DN, is an expression of the particular knowledge it conveys to its community of users as an information resource. Notice that different information resources may share a very large set of keywords and documents. However, these are organized differently in each resource, leading to different associative semantics. Indeed, each resource is tailored to a particular community of users, with a distinct history of utilization and deployment of information. For instance, the same keywords will be related to different sets of documents in distinct resources, thus resulting in different distances for the same pairs of keywords. Therefore, we refer to the relational information, or associative semantics, of each information resource as a *Knowledge Context* (Rocha 2001b). We do not mean to imply that information resources possess cognitive abilities. Rather, we note that the way documents are organized in information resources is an expression of the knowledge traded by their communities of users. Documents and keywords are only tokens of the knowledge that is ultimately expressed in the brains of users. A knowledge context simply mirrors some of the collective knowledge relations and distinctions shared by a community of users. The distance graphs which relate elements of DN define an associative semantics. They convey how strongly associated pairs of elements in the specific network are.

More specifically, we characterize an information resource *R* by a structure named Knowledge Context:

$$KN_R = \{\mathcal{X}, \mathcal{R}, d\} \quad (3)$$

Where $\mathcal{X}$ is a set of available sets of elements $X_i$, e.g. $\mathcal{X} = \{K, M, U\}$, where *K* is a set of keywords, *M* a set of documents, and *U* a set of users. $\mathcal{R}$ is a set of available relations amongst the sets in $\mathcal{X}$, e.g. $\mathcal{R} = \{\mathbf{C}(M, M), \mathbf{A}(K, M)\}$, where **C** denotes a citation relation between the elements of the set of documents, and **A** a semantic relation between documents and keywords, such as the keyword-document matrix defined in section 2.1. Finally, *d* is a set of distance functions applicable to some subset of relations in $\mathcal{R}$, e.g. $d = \{d_k\}$, where $d_k$ is a distance between keywords such as the one defined by formula (2). The application of these distance functions results on distance graphs *D* whose vertices are elements from the sets in $\mathcal{X}$.

In our architecture of recommendation (Rocha 2001b), users are themselves characterized as information resources, where $\mathcal{X}$ may contain, among other application-specific elements, the sets of documents previously retrieved by the user and their associated keywords. Ultimately, what feeds recommendation algorithms, are the distance functions *d* of knowledge contexts. In this article, we deal in particular with the metric behavior of such functions. We discuss below how an analysis of the metric behavior of distance graphs extracted from DN, allows us to produce appropriate recommendations, and also to better understand the quality of the knowledge stored in information resources. We note that the metric analysis detailed below is not tied to our particular view of recommendation systems, which we discuss in section 9.

## 3. Semi-Metric Behavior

The distance graph $D$ obtained from applying distance function $d$ (formula 2) to the KSP proximity relation obtained from formula (1), is not Euclidean because, for a pair of keywords $k_1$ and $k_2$, the triangle inequality may be violated: $d(k_1, k_2) \leq d(k_1, k_3) + d(k_3, k_2)$ for some keyword $k_3$. This means that the shortest distance between two keywords in $D$ may not be the direct link but rather an indirect pathway[4]. Such distance functions are referred to as semi-metrics (Galvin & Shore 1991).

Naturally, the distance graphs obtained from applying a distance function such as $d$ can be made Euclidean. If we compute the transitive closure of the proximity relation derived from formula (1), we would obtain a similarity relation on the set of keywords. The application of distance $d$ (formula 2) to a similarity graph would result in a metric distance graph. Indeed, it is very common in the analysis of social or document networks to impose a metric requirement on the distance graphs generated. The purpose of this article is to show that loosening the metric requirement, results in a methodology capable of identifying important characteristics of DN – which we loose with metric distance graphs.

Most ideas are born out of anecdotal, often personal, evidence. The one put forward here is no exception. It arose from questioning what could one infer from the semi-metric behavior of the distance graphs calculated from DN. Given a distance function, what can we say about a pair of highly semi-metric elements from a finite set? And what can we say about the resulting distance graph, from the pairs of highly semi-metric pairs it contains? To construct an intuition to answer these questions, one needs to deal with very

---

[4]Given that most social and knowledge-derived networks possess Small-World behavior (Watts 1999), we expect that vertices which tend to be clustered in a local neighborhood of related vertices, show large distances to vertices in other clusters. But because of the existence of "gateway" vertices relating vertices in different clusters (the small-world phenomenon), smaller indirect distances between vertices in distinct clusters, through these "gateway" vertices, are to be expected.

familiar examples. In this case, the author could think of no DN more familiar than the set of books cited by his own dissertation (Rocha 1997a)! A database similar (but much smaller) to the one used by ARP contains the relevant information. This database contains about 150 books, each indexed by the respective Library of Congress Keywords, for example:

Kearfott, R. Baker and Vladik Kreinovich (Editors) (1996). *Applications of Interval Computations*. Kluwer. Keywords: Optimization algorithms, Fuzzy logic, Uncertainty, Mathematics, Reliable Computation, Interval Computation.

Table 2: Distance function for 5 keywords in the dissertation database

|  | Adaptive Systems | Evolution | Modeling systems | Complex Systems | Social Systems |
|---|---|---|---|---|---|
| Adaptive Systems | 0.00 | 3.89 | 12.00 | 10.33 | 16.00 |
| Evolution | 3.89 | 0.00 | 21.50 | 4.22 | 35.00 |
| Modeling Systems | 12.00 | 21.50 | 0.00 | 5.75 | 10.00 |
| Complex Systems | 10.33 | 4.22 | 5.75 | 0.00 | 19.00 |
| Social Systems | 16.00 | 35.00 | 10.00 | 19.00 | 0.00 |

From this database, 86 keywords were extracted. A distance graph $D$ was calculated using function $d$ according to formula (2). Table 2 shows the values of the edges of this graph $d(k_1, k_2)$ for 5 of the keywords. One needs to note that this distance graph is obtained from the relations extracted from a very particular set of documents (in this case 150 books). Therefore, one should not expect the distance values to represent a universally accepted thesaurus or the associations one would anticipate from common sense semantics. Indeed, this kind of distance function is used to characterize particular information resources, as discussed in section 2. In this case, the distance graph on the set of keywords denotes only the derived associative semantics from the set of books cited in the dissertation.

To obtain the shortest distance between vertices (keywords) of $D$, we used a (+, min) matrix composition of $D$ until closure is achieved – ote that traditional algebraic matrix composition is (*, +). Table 3 shows the shortest distances for the same 5 keywords. We see for instance that the shortest indirect distance between MODELING SYSTEMS and EVOLUTION is 9.97, whereas the direct distance is 21.5. This means that the distance between the keyword pair MODELING SYSTEMS-EVOLUTION is semi-metric. This is not the case of the metric pair ADAPTIVE SYSTEMS-EVOLUTION, for which the shortest distance is the direct one.

Table 3: Shortest distance for 5 keywords in the dissertation database (semi-metric pairs shown in italics).

|  | Adaptive Systems | Evolution | Modeling systems | Complex Systems | Social Systems |
|---|---|---|---|---|---|
| Adaptive Systems | 0.00 | 3.89 | 12.00 | *8.11* | 16.00 |
| Evolution | 3.89 | 0.00 | *9.97* | 4.22 | *19.89* |
| Modeling Systems | 12.00 | *9.97* | 0.00 | 5.75 | 10.00 |
| Complex Systems | *8.11* | 4.22 | 5.75 | 0.00 | *15.75* |
| Social Systems | 16.00 | *19.89* | 10.00 | *15.75* | 0.00 |

## 4. Characterizing Semi-metric Behavior

Clearly, semi-metric behavior is a question of degree. For some pairs of vertices the indirect distance provides a much shorter short-cut, a larger reduction of distance, than for others. One way to capture this property of pairs of semi-metric vertices (keywords) is to compute a *semi-metric ratio*:

$$s(k_i, k_j) = \frac{d_{direct}(k_i, k_j)}{d_{shortest}(k_i, k_j)} \qquad (4)$$

where $d_{shortest}$ is the shortest distance between the keyword pair. $s$ is positive and $\geq 1$ for semi-metric pairs, where $d_{shortest}$ in an indirect distance between the two keywords. In our example, $s$(MODELING SYSTEMS, EVOLUTION) = 21.5/9.97 = 2.157. This ratio is important to discover semi-metric behavior necessary for our analysis as discussed below, but given that larger graphs tend to show a much larger spread of distance, $s$ tends to increase with the number of vertices of $D$. Therefore, to be able to compare semi-metric behavior between different DN and their respective different sets of keywords, a *relative semi-metric ratio* is used:

$$rs(k_i, k_j) = \frac{d_{direct}(k_i, k_j) - d_{shortest}(k_i, k_j)}{d_{max} - d_{min}} \quad (5)$$

*rs* compares the semi-metric distance reduction to the maximum possible distance reduction in graph *D*. $d_{max}$ is the largest distance in the graph, and $d_{min} = 0$ is the shortest distance in the entire graph. This ratio varies between 0 and 1 for semi-metric pairs, and it is negative for metric pairs.

Often, the direct distance between two keywords is ∞ because they do not index any documents in common. As a result, *s* and *rs* are also ∞ for these cases. Thus, *s* and *rs* are not capable of discerning semi-metric behavior for pairs that do not have an initial finite direct distance. To detect relevant instances of this infinite semi-metric reduction, we define the *below average ratio*:

$$b(k_i, k_j) = \frac{\overline{d_{k_i}}}{d_{shortest}(k_i, k_j)} \quad (6)$$

where $\overline{d_{k_i}}$ represents the average direct distance from $k_i$ to all $k_k$ such that $d_{direct}(k_i, k_k) \geq 0$. *b* is only applied to semi-metric pairs of keywords ($k_i$, $k_j$) where $d_{shortest}(k_i, k_j) < d_{direct}(k_i, k_j)$ and it measures how much the shortest indirect distance between $k_i$ and $k_j$ falls below the average distance of all keywords $k_k$ directly associated with keyword $k_i$. Note that $b(k_i, k_j) \neq b(k_j, k_i)$. Of course, *b* can also be applied to pairs with finite semi-metric reduction. $b > 1$ denotes a below average distance reduction.

# 5. Analysis of a Collection of Documents: The Global Interests of the Collector

The three semi-metric ratios were applied to graph *D* of the dissertation database. Table 4 lists the top 5 pairs for semi-metric ratio *s*. If we rank pairs for the relative semi-metric ratio *rs*, there is a slight reordering of the top as the pair EVOLUTION-DNA drops to rank 11 and the pair LIFE-COGNITION to 6[th], while the pair EVOLUTION-CONTROL rises to rank 3 (from 6[th]) and the pair EVOLUTION-INFORMATION THEORY rises to 5[th] (from 20[th]).

Table 4: Semi-metric pairs with highest *s* in dissertation database.

| $(k_i, k_j)$ | $s(k_i, k_j)$ | $rs(k_i, k_j)$ |
|---|---|---|
| ADAPTIVE SYSTEMS-COGNITION | 6.39 | 0.84 |
| EVOLUTION-CONSTRUCTIVISM | 5.00 | 0.76 |
| EVOLUTION-PSYCHOLOGY | 5.00 | 0.73 |
| EVOLUTION-DNA | 4.69 | 0.64 |
| LIFE-COGNITION | 4.55 | 0.66 |

What is most interesting about these results is that these pairs denote the original contributions that were offered by the dissertation! Indeed, the dissertation was about using ideas and methodologies from Complex Adaptive Systems, Evolutionary Systems, and Artificial Life and apply them to Artificial Intelligence and Cognitive Science. In particular, the mathematical models (from Psychology) of cognitive categories were expanded using evolutionary ideas, by drawing an analogy with the symbolic characteristics of DNA. Furthermore, this framework was named Evolutionary Constructivism, a term that did not exist previously, but draws both from Evolutionary Theory and the Philosophy of Constructivism in Cognitive Science and Systems Theory.

To understand these results, we need to remember that the distance function *d* is derived from the finite set of books used in the dissertation. A high degree of semi-metricity for a keyword pair means that this pair of

keywords co-indexes very few of the books in the database, but that there exists a strong indirect association between the two keywords via some indirect path, whose short distances require the existence of many books co-indexed by the keyword pairs in the edges of the indirect pathway. Thus, a keyword pair with high values of semi-metric ratios, implies a strong keyword association that is a global property of the specific collection of documents, but not one identifiable in many included documents, and rather constructed from an indirect series of strongly related documents. In other words, the highly semi-metric pairs represent associations that are *latent* in this specific collection of documents – novel associations "begging to be made". Indeed, the two pairs (EVOLUTION-CONTROL and EVOLUTION-INFORMATION THEORY) ranked in the top 5 for the relative semi-metric ratio, identify two associations that are certainly implied by the collection of books (given its large subsets of Cybernetics and Information Theory books), but which were not dealt with in this dissertation – offering some topics for other dissertations!

We can see how the semi-metric analysis of a set of documents associated with a person, say the set of documents retrieved by a user of some information resource, can reveal a set of implied interests of the user, which she has not been able to satiate with individual documents (she may not even be aware of the need). These refer to those interests which are implied by the global associative semantics, but not by previously retrieved (or collected) individual documents. In this sense, semi-metric pairs discover a demand for novel documents to fill a gap implied by the overall DN. If we had used a metric distance function, this specific need would go unnoticed. Finally, the ability to discover an implied demand for documents with this semi-metric analysis is clearly useful for recommendation systems. We detail the development of such systems in a separate article. Here, we want to strengthen our evidence of the utility of semi-metric analysis by applying it to several other larger DN, and proposing a classification of DN according to semi-metric behavior.

## 6. Analysis of Larger Datasets: Latent Trends

The anecdotal analysis of the author's dissertation database served the purpose of creating an intuition of what semi-metric behavior may mean for

DN, but we also applied it to other more "subject-independent" datasets . The same semi-metric behavior ratios were used to study the ARP dataset describe above. A distance graph $D$ was calculated using function $d$ (formula 2) for the set of the 500 most common keywords. The semi-metric ratios (formulas 4 to 6) were then calculated for all edges of $D$ (keyword pairs). Table 5 shows the 5 keyword pairs with highest values of $s$.

Table 5: Semi-metric pairs with highest $s$ in ARP dataset.

| $(k_i,k_j)$ | $s(k_i,k_j)$ | $rs(k_i,k_j)$ |
|---|---|---|
| LEUKEMIA-MYOCARDI | 272.20 | 0.4981 |
| HORMON-THIN | 214.08 | 0.9953 |
| CARE-EXCIT | 213.59 | 0.9953 |
| GENE-EQUAT | 205.76 | 0.9951 |
| FILM-TRANSCRIPT | 204.51 | 0.9951 |

To analyze these results, again, one must remember the original collection of documents, in this case, all the scientific articles published in journals indexed by *ISI* in *SciSearch* between the years of 1996 to 1999. An edge of $D$ with high values of semi-metric ratios $r$ and $rs$, implies that while very few articles are co-indexed by the respective keyword pair, a series of articles exists which creates an indirect path in $D$ between these two keywords. To obtain a large semi-metric ratio, it is necessary that all edges in the shortest indirect path be defined by short direct distances, which in turn require the existence of many articles co-indexed by both keywords in every edge of this path. Thus, a highly semi-metric keyword association between two keywords implies that very few documents are co-indexed by the keyword pair, but that there are sets of documents indirectly supporting the pair.

The existence of indirect support for a pair of keywords (particularly in scientific databases) may identify a trend that can be expected to be picked up. While it is hard to understand all associations identified in a dataset such as ARP containing so many different topics, at least one association is observed in the data set which is meaningful to the author. The high semi-metricity of the GENE-EQUAT[5] pair may be a result of the trend observed in the late 1990's towards computational and mathematical biology, as molecular biology started to move into a post-genome bioinformatics mode (Kanehisa 2000).

Indeed, the analysis of the keyword pairs with infinite semi-metric reduction characterized by a high below average ratio $b$, seems to give further evidence for this claim, as the highest values of $b$ are observed for the pairs EQUAT-MESSENGERRNA, EQUAT-TRANSCRIPT, and EQUAT-GENE-EXPRESS. These pairs associate the keyword *Equation* with keywords that describe the chief technology that enabled the greatest advances in bioinformatics in the late 1990's and today: the *Gene Expression* Arrays that allow the rapid measurement in parallel of *messenger RNA transcribed* from DNA in the cell (the process of gene expression). Ratio $b$ seems to be useful for larger datasets not collected by a single author. As discussed in more detail in section 7, such large datasets are very incomplete in the sense that many potentially semantically associated keyword pairs do not co-index a single document, resulting in an infinite distance for those pairs. Ratio $b$ picks associations between those pairs of keywords that do not co-index a single document but that are strongly implied by the overall collection.

In the case of the ARP database, which contains almost 3 million documents (see section 2.1) collected between 1996 and 1999, not a single document was co-indexed by the keywords in the pairs picked up with high values of $b$ (e.g. *Equation* and *Gene Expression*). But the entire collection of documents strongly implied such keyword associations, identifying a *latent* relationship in the literature. We can interpret a latent association between two keywords as evidence that a direct association may be instantiated later by the appearance of documents co-indexed by both keywords, in other words, it identifies a plausible trend. Indeed, we note that in 2001, we observed that in the same collection of journals used by ISI and SciSearch, a very small set of papers started appearing which are co-indexed by both keywords *Equation* and *Gene Expression*, for instance:

---

[5]Notice that ARP keywords are stemmed to group different constructions of the same term: e.g. Equation and Equations.

Pasto M et al [2001]. "Metabolic activity of the external intercostal muscle of patients with COPD". *Archivos De Bronconeumologia*, v. 37(#3), pp. 108-114, Mar 2001

We expect to observe more articles using this pair of keywords, as the publication of papers in scientific journals typically lags 1 or 2 years from the submission date. In any case, the semi-metric ratios defined in section 4 are useful to identify latent associations implied by a collection of documents, and to give a measure of strength for this latent, semi-metric behavior. Below we use them also to characterize different DN, including known public datasets.

We would like to note that in information retrieval, the term latent is associated with a particular technique known as *Latent Semantic Analysis* (LSA) (Landauer, P.W.Foltz, & D.Laham 1998) or *Latent Semantic Indexing* (Berry, DUMAIS, & Obrien 1995). This technique uses Singular Value Decomposition (related to Principal Component Analysis) to group keywords which are semantically associated directly or indirectly in a collection of documents. In this sense, our usage of the term latent is similar to LSA. But our semi-metric analysis is proposed here both as a means to identify those <u>specific</u> pairs of keywords which are *most* latent (useful for recommendation), and as discussed in section 7, to characterize types of networks (useful to advance our knowledge of networks).

## 7. Characterizing Document Networks

The behavior of the three semi-metric ratios above (formulas 4 to 6) can also be used characterize the type of DN we encounter. We expect different DN to possess different semi-metric behavior. As discussed in section 2.3, we assume that the way keywords are semantically associated in information resources is an expression of the knowledge traded by their communities of users. Thus, we expect semi-metric behavior to allow us to better understand how knowledge is stored in DN, and furthermore determine types of DN.

## 7.1 Additional Document Networks

To investigate this hypothesis, in addition to the dissertation and ARP databases, we have applied this study to other DN such as[6]:

1.  ***Distance graph built from web-site Hyperlink Structure*** (PCPStruct). We used a structural proximity measure computed by Bollen (Bollen 2001) for all pairs of the 423 web pages of the *Principia Cybernetica Project* (PCP)Web site (http://pespmc1.vub.ac.be/), which its editors deemed most important. This web site is a collection of dictionary-like definitions about Systems Research topics; each of these 423 web pages is associated with a specific concept (e.g. "Adaptive Systems"). These web pages/concepts were taken as vertices of a non-directed graph, whose edges are weighted with a value computed by a structural proximity measure very similar to formula (1): $P_{struc} = \max(P^{in}, P^{out})$. $P^{in}$ is an inwards proximity where two web pages are considered near if they tend to be linked from many of the same pages, formally it is the probability that two web pages are both linked from a page that links to one of them. $P^{out}$ is an outwards proximity where two web pages are considered near if they tend to link to the same pages, formally it is the probability that two web pages tend to link to the same page that one of them links to(see (Rocha & Bollen 2001) for details). We then normalized all $P_{struc}$ values linearly against the highest value. Finally, using formula (2), we obtained a distance graph $D$ for the set of web pages.
2.  ***Distance graph built and adapted from web-site collective usage*** (PCPAdap). Based on the methodology of (Bollen & Heylighen 1998), Bollen (Bollen, Vandesompel, & Rocha 1999) utilized user paths derived from the web server logs of the *Principia Cybernetica Project* (PCP)Web also used in 1. These web pages/concepts were taken as vertices of a directed graph, whose edges are weighted with a value computed from three reinforcement rules: *frequency* (reinforces the edge a → b every time it is traversed by a user), *symmetry* (reinforces the edge b → a every time a → b is traversed), *transitivity* (reinforces the edge a → c every time the path a → b → c is traversed)[7]. For the present work, we constructed a non-directed graph from this user-adapted network by taking the maximum of both directed links between two vertices. We then normalized the weights of this graph linearly against its highest value, which we took as a measure of proximity between vertices. Finally, using formula (2), we obtained a distance graph *D*.

---

[6]The additional DN used here were produced by Johan Bollen for the Active Recommendation Project. The author wishes to thank him for making them available.

3. ***Distance graph built and adapted from usage of Journal titles in LANL's research digital library*** (ISSN). Similarly, Bollen (Bollen & Vandesompel 2000) utilized user paths derived from the weblogs of the Los Alamos National Laboratory Research Digital Library (http://lib-www.lanl.gov/) to adapt a network of 472 Research Journal Titles (e.g. "Communications of the ACM" and "BioSystems"), identified by their ISSN. Using the same methodology used for PCPAdapt (point 2 above), we obtained a distance graph on the set of these journal titles.

4. ***Distance graph built from free word association norms*** (Word Norm) Nelson et al (Nelson, McEvoy, & Schreiber 1998) have computed tables of associations between pairs of words from free association experiments with more than 6000 subjects. These tables are in effect directed graphs between words, whose weights are taken to characterize the semantic proximity between words as understood by the population of subjects. These kinds of norms can be seen as a measurement of the associative semantics of a population. In the present work, we used a subset of 150 words from this dataset (of about 5000 words). Our 150 words were the same used by Bollen and Heylighen (Bollen & Heylighen 1998) in their hypertext experiments, described as the 150 most common English nouns, (words such as "art", "car", "face"). Similarly to the cases above, we constructed a non-directed graph by taking the maximum of both directed links between two words. We then normalized the weights of this graph linearly against its highest value, which we took as a measure of proximity between vertices. Finally, using formula (2), we obtained a distance graph *D*.
5. ***Random distance graphs***.
    a. *Uniform*. We computed 50 proximity graphs of 150 vertices, and 20 of 500 vertices, whose random weights are uniformly distributed in the unit interval. Using formula (2), we derived the respective distance graphs.
    b. *Exponential*. An analysis of the weights of the PCPAdapt, ISSN, and Dissertation proximity graphs reveals that they fit an exponential random distribution[8] ($\lambda$ = 15.3; 18.8; and 10.15 respectively). We computed exponential proximity graphs with $\lambda$ = 5, 10, 15, 20, 30, 40, and 100. For each $\lambda$, we produced 10 proximity graphs of 150 vertices, and 5 of 500 vertices. Using formula (2), we derived the respective distance graphs.
    c. *Hyper-exponential*. An analysis of the weights of the PCPStruct, ARP, and Word Norm proximity graphs reveals that they fit a hyper-exponential random distribution[9] ($\lambda_1$ = 8.3, $\lambda_2$ = 39.2, $p$ = 0.38; $\lambda_1$ = 8.8, $\lambda_2$ = 37.8, $p$ = 0.11; and $\lambda_1$ = 7.6, $\lambda_2$ = 33.3, $p$ = 0.5 respectively). We computed random hyper-exponential proximity graphs with similar parameters: $\lambda_1$ = 8, $\lambda_2$ = 39, $p$ = 0.4; $\lambda_1$ = 9, $\lambda_2$ = 38, $p$ = 0.1; and $\lambda_1$ = 8, $\lambda_2$ = 33, $p$ = 0.5 . For each of these two cases, we produced 10 proximity graphs of 150 vertices, and 5 of 500 vertices. Using formula (2), we derived the respective distance graphs.

---

[7]For more details of this adaptive hypertext mechanism see (Bollen & Heylighen 1998, Bollen, Vandesompel, & Rocha 1999), or (Rocha & Bollen 2001).

## 7.2 Semi-metric Behavior: Measuring Associative Semantics

Figure 1 depicts the percentages of pairs with semi-metric behavior described by ratios *rs* (formula 5) and *b* (formula 6) for all non-random distance graphs described above. Table 6 summarizes all the values. From this figure alone we can observe that the Word Norm distance graph shows very small semi-metric behavior: only 1% of all word pairs have $rs > 0.0$, and only 10% have $b > 1.0$. We would expect this behavior since, given the large pool of subjects in the Word Norm experiments, the appropriate distance between pairs of the most common English words has been directly established. We can say that the Word Norm distance graph captures most of the common sense word associations in a population; it is a fairly *complete* measurement of its semantics.

---

[8] The exponential probability distribution function is $F(t) = \lambda e^{-\lambda t}$, $t \geq 0$, $\lambda > 0$.

[9] The hyper-exponential probability distribution function is $F(t) = \lambda_1 p e^{-\lambda_1 t} + \lambda_2 (1-p) e^{-\lambda_2 t}$, $t \geq 0$, $\lambda_1, \lambda_2 > 0$.

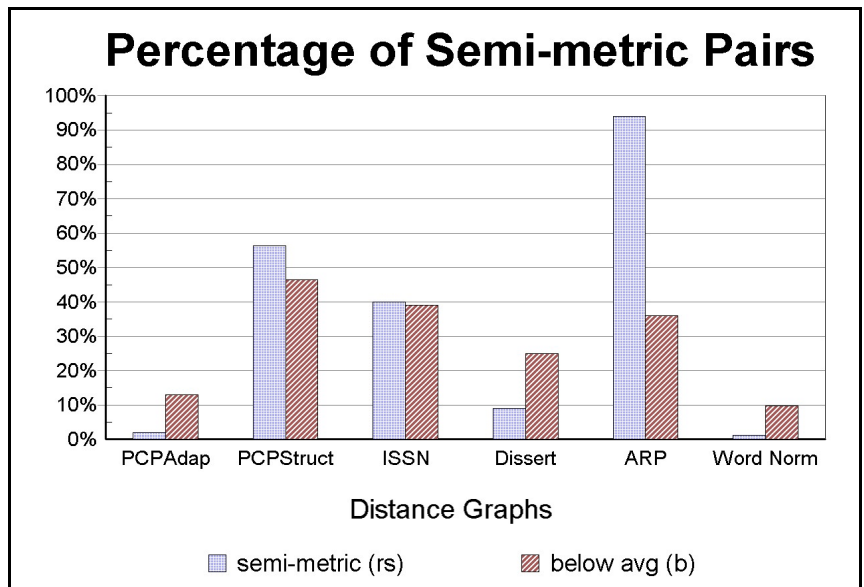

**Figure 1**: Semi-metric behavior of the several non-random distance graphs.

The ARP distance graph on the other hand, has been built from automatic analysis of keywords present in a large set of scientific documents from several fields of inquiry (e.g. Biology, Physics, Chemistry). In this dataset, we expect the same keyword to have distinct meanings for distinct communities of authors (polysemy) and the same concept to be described by several keywords (synonomy). Furthermore, when keywords co-index a document, unlike the Word Norm experiments, authors and/or editors are not expected to list possibly associated keywords. Thus the ARP distance graph is very far from capturing all the possible, relevant, associations between pairs of keywords, it indeed provides only a very incomplete measurement of the semantics of the population of its authors. This is apparent from the percentage of semi-metric pairs observed: 94% of all

Table 6: Percentage of semi-metric pairs for ratios *rs* and *b* for all distance graphs. Values for random cases are averages, standard deviation also shown.

|  | $N$ | %*rs* | stdv %*rs* | %*b* | stdv %*b* |
|---|---|---|---|---|---|
| ARP | 500 | 94 |  | 36 |  |
| Dissert | 86 | 9 |  | 25 |  |
| PCPAdap | 423 | 2 |  | 13 |  |
| PCPStruct | 423 | 56 |  | 47 |  |
| ISSN | 472 | 40 |  | 39 |  |
| Word Norm | 150 | 1 |  | 10 |  |
| Uniform | 150 | 96.3223 | 0.2166 | 14.7703 | 0.5616 |
|  | 500 | 98.6466 | 0.0307 | 12.3738 | 0.2558 |
| Exp $\lambda=5$ | 150 | 97.939 | 0.2699 | 15.4505 | 0.4949 |
|  | 500 | 99.2856 | 0.0093 | 13.2108 | 0.3351 |
| Exp $\lambda=10$ | 150 | 87.9974 | 0.4723 | 15.208 | 0.6997 |
|  | 500 | 93.6482 | 0.0444 | 14.3884 | 0.0757 |
| Exp $\lambda=15$ | 150 | 85.029 | 0.4908 | 16.0492 | 0.5618 |
|  | 500 | 90.5042 | 0.0596 | 14.784 | 0.0972 |
| Exp $\lambda=20$ | 150 | 83.3281 | 0.3757 | 16.6791 | 0.7992 |
|  | 500 | 89.0098 | 0.1100 | 15.4748 | 0.2250 |
| Exp $\lambda=30$ | 150 | 82.2659 | 0.4815 | 17.3846 | 0.5559 |
|  | 500 | 87.8064 | 0.0968 | 16.1488 | 0.1528 |
| Exp $\lambda=40$ | 150 | 81.6992 | 0.5263 | 18.0295 | 0.4917 |
|  | 500 | 87.1862 | 0.1472 | 16.8334 | 0.2021 |
| Exp $\lambda=100$ | 150 | 80.1878 | 0.3540 | 20.3974 | 0.4448 |
|  | 500 | 85.7592 | 0.1337 | 19.618 | 0.1233 |
| HyperExp (8, 39, 0.4) | 150 | 92.1146 | 0.21 | 16.655 | 1.0188 |
|  | 00 | 96.1634 | 0.451 | 16.0852 | 0.2362 |
| HyperExp (8, 33, 0.5) | 150 | 97.939 | 0.3230 | 15.4505 | 0.5107 |
|  | 500 | 95.9424 | 0.1351 | 15.342 | 0.1588 |
| HyperExp (9, 38, 0.1) | 150 | 91.5947 | 0.3893 | 17.6178 | 0.4240 |
|  | 500 | 96.3464 | 0.0661 | 16.7052 | 0.0991 |

keyword pairs have $rs > 0.0$ (meaning that there is a shorter indirect distance for most pairs), and 36% have $b > 1.0$.

The other datasets also observe an expected semi-metric behavior. The PCPAdap distance graph ($rs$: 2% and $b$: 13%) behaves similarly to the Word Norm case, which shows that Bollen's adaptive hypertext algorithm succeeded in capturing the semantics of its user community in a fairly complete manner. Especially when we contrast it to the PCPStruct distance graph ($rs$: 56% and $b$: 47%), which contains many more pairs of web pages/concepts with shorter indirect distances. The distance graph of PCPStruct was built exclusively from the hyperlink structure of the web site as designed by its authors. Because we observe a large percentage of semi-metric pairs, we can conclude that many associations between web pages are not explicitly made by the web site authors, but implied by the overall hyperlink structure. Using Bollen's algorithm on this same web site, which integrates the traversal paths from the user population with a transitivity rule, brings about a substantial reduction of indirect associations. In other words, the population of users whose traversal behavior was integrated by Bollen's algorithm, identified most of the relevant associations between web pages.

However, the same adaptive hypertext algorithm, was not as successful in deriving a complete measurement of the semantic associations between journal titles in the ISSN distance graph ($rs$: 40% and $b$: 39%). Several reasons for this can be found. The user community used to adapt the PCP web site is more thematically coherent. PCP is a web site dedicated to the study of Systems Research, thus, its community of users functions in the same research universe. In contrast, the user community used to adapt the ISSN distance graph is the population of scientists and engineers at the Los Alamos National Laboratory, which contains a rather diverse pool of people from such fields as Physics, Biology, Material Science, Computer Science, etc. Indeed, the 472 journals in this network cover all areas of science and technology. Thus, such a heterogeneous community may not possess a complete associative semantics to begin with. By this we mean that distinct communities fail to see associations outside of their usual set of journals and terminology. As shown in section 6, in such heterogeneous DN such as ARP and ISSN, many relevant associations are not explicitly observed in individual documents or by users of the dabatases, but rather implied by the network.

Furthermore, Bollen's (Bollen 2001) methodology to recover user paths from user logs may not be as appropriate for the ISSN network as it was for the PCP case. Whereas PCP users were browsing a web site, ISSN users were retrieving documents from a digital library. The first were more prone to pursue links thus creating longer paths which can be used by the transitivity rule of Bollen's algorithm to discover indirect associations. The users of a research library tend to be looking for specific papers from references or retrieving documents from keyword searches. This fact could perhaps be alleviated by collecting web logs for a longer period of time.

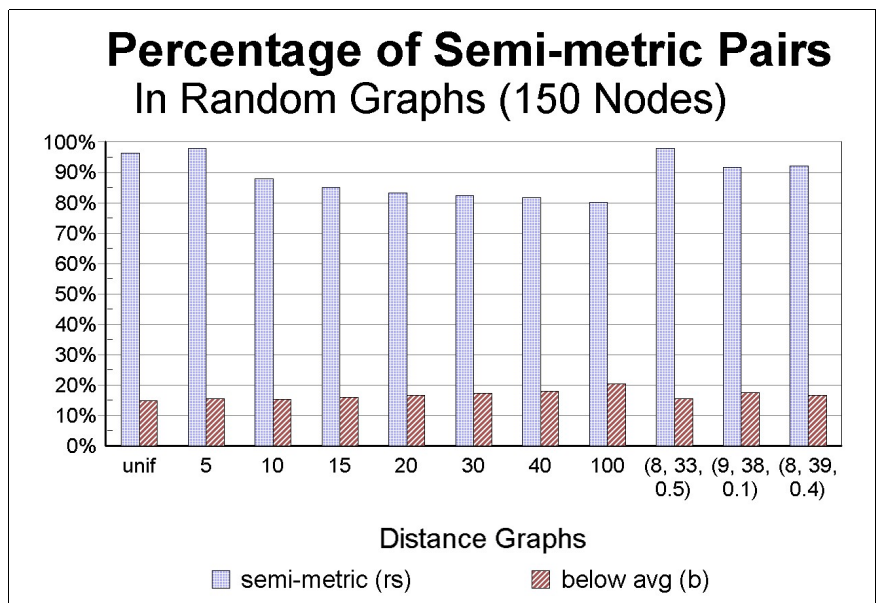

**Figure 2**: Average values of semi-metric ratios for random distance graphs with 150 vertices. Leftmost distance graph is for the uniform random case. Rightmost three graphs refer to the hyper-exponential distance graphs discussed above, with ($\lambda_1$, $\lambda_2$, $p$). All other distance graphs refer to the various exponential cases identified by $\lambda$.

The dissertation distance graph (*rs*: 9% and *b*: 25%) also shows small semi-metric behavior, though larger than the Word Norm and PCP cases. This is also expected since it was extracted from the keywords of all the books referred to in a dissertation. Such a collection of books is necessarily fairly related in order for the dissertation to be a coherent piece. We have discussed in section 5 that the semi-metric pairs discovered in this distance graph revealed what was novel in the dissertation. The low numbers of semi-metric pairs reveal that the distance graph is also a fairly complete measurement of the associative semantics implied by the dissertation. This supports the usage of distance functions in the knowledge context structure (section 2.3) to characterize the semantic interests of a user in recommendation systems, since a single user tends to collect fairly related documents.

Figure 2 depicts the average percentage of semi-metric pairs in the various random distance graphs with 150 vertices. Table 6 also summarizes the values for random distance graphs with 150 and 500 vertices, and also indicates the standard deviation of these values. In all random cases, we observe high percentages of semi-metric pairs (though not very large values of below average measure *b*). This shows that in all types of random graphs, most pairs of vertices possess some smaller indirect distance pathway. By contrasting figures 1 and 2 (or Table 6), we notice immediately that the semi-metric behavior of the random distance graphs is similar to that of the ARP distance graph. Nevertheless, we can show that the semi-metric behavior of random graphs, where we do not expect a real semantics to exist at all, is quite different from the behavior of graphs such as ARP where some semantics does exist, even if fractional or incomplete. To better distinguish between random distance graphs and distance graphs such as ARP produced from large collections of documents, we study their semi-metric behavior in more detail in the next subsection.

We end this section by concluding that we compiled and discussed evidence that the percentage of pairs with semi-metric behavior in a distance graph can be a good indicator of how well such graph captures the associative semantics of a DN from where it is derived. On one extreme we have Word Norms, a complete semantic measurement of common words, with low percentages of semi-metric pairs, and on the other, DN such as ARP, a very fractional measurement of keyword semantics, with high percentages of semi-metric pairs. Next we discuss another dimension of semi-metric behavior, and propose a means to classify the nature of the associative semantics entailed in DN.

## 7.3. The behavior of the most semi-metric-pairs: Strength of Latent Associations

The distance graphs analyzed are of different sizes, from 86 keywords (3655 pairs) in the dissertation database, to 500 in the ARP case (124750 pairs). To compare the semi-metric behavior of all these DN of different sizes, we selected 1% of all pairs with highest *rs* in the respective distance graph *D*. In other words, we ranked all pairs according to *rs*, and then selected the top 1%[10]. We chose 1% of all pairs because this is the percentage of semi-metric pairs found in the Word Norm dataset, which displays the smallest number of semi-metric pairs. The Word Norm clearly functions in this analysis as a benchmark since it contains as good an example of a measurement of a population's associative semantics as one can get. This way, we can compare the semi-metric behavior of the other distance graphs against a complete picture of the behavior of a very complete associative semantics. Table 7 lists the numbers of pairs that 1% represents for each dataset.

Table 7: Number of Pairs in top 1% pairs

| ARP | Dissertation | PCP | ISSN | Random 150 | Random 500 | Word Norms |
|---|---|---|---|---|---|---|
| 1248 | 37 | 893 | 1112 | 112 | 1248 | 112 |

To obtain a graphic model of semi-metric behavior, we further need to normalize the number of pairs. Thus, for each ranked list of top 1% pairs, we sampled 100 equally spaced (in rank) pairs, except for the dissertation dataset where we expanded the existing 37 pairs to 100 using linear interpolation of the values of *rs*. We denote this ranked, sampled (or expanded) set of pairs as $TRS^{1\%}$. Figure 3 depicts the values of *rs* for the pairs in $TRS^{1\%}$ for each distance graph. Each curve depicts the normalized

---

[10]Not 1% of the set of semi-metric pairs, but 1% of all pairs in *D*.

semi-metric behavior of the top 1% most semi-metric pairs for each distance graph.

One thing we immediately notice in figure 3 is that even though both the ARP and uniform random distance networks produce large percentages of semi-metric pairs (figures 1 and 2), the behavior of *rs* is quite different in their $TRS^{1\%}$ sets. The uniform random graphs, except for a very small number of pairs, show very low values of *rs* which quickly decay to close to zero. In fact, almost all of the many semi-metric pairs in uniform random graphs possess a very small value of *rs*, whereas in the ARP case, semi-metric pairs possess a substantially higher value of *rs*.

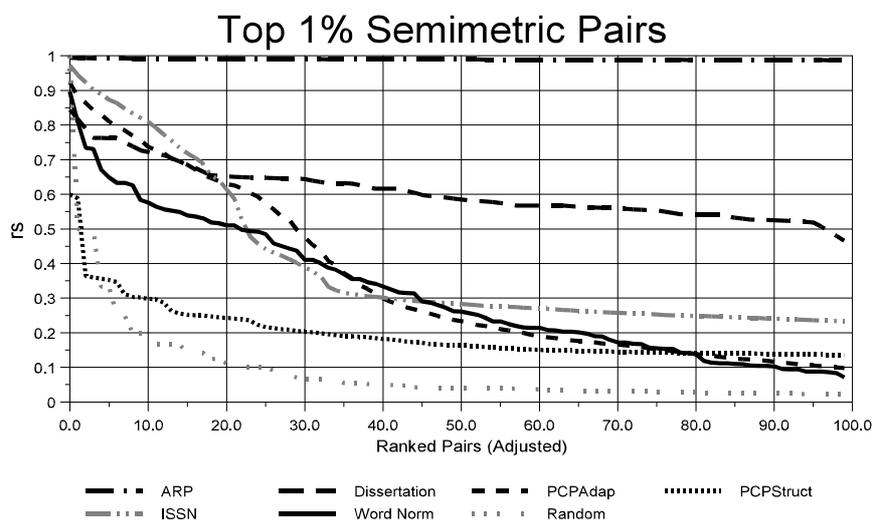

**Figure 3**: Top 1% semi-metric pairs sampled or expanded to a series of 100 points. Random depicts a uniform random case

This shows, as expected, that vertices in a uniform random graph are not related in a semantically coherent manner. For most pairs of vertices, there tends to exist a shorter indirect distance (semi-metric path), but which nonetheless tends not to be much shorter than the direct path, because the distances of the edges in the indirect paths are also computed from a uniformly random proximity distribution, and thus tend not to reduce the distance by much. The longer the indirect path between a pair of vertices, the smaller the probability of it offering a shorter path than the direct distance between the pair[11]. ARP, on the other hand, is clearly not random. Most of its keyword pairs do have shorter indirect paths, but many of these offer substantial distance reductions. This means that there exist plenty of strong latent associations. Both the ARP and uniform random distance graphs are incomplete as a representation of associative semantics, since both have large percentages of semi-metric pairs (figures 1 and 2), but most of the semi-metric associations in the uniform random distance graphs are very weak, whereas the ARP distance graph contains many strong ones (figure 3).

## 7.4 Quantifying Latent Associations in Document Networks

Using Semi-metric BehaviorWe need now to compare the semi-metric behavior of real distance graphs such as ARP with the more realistic random distance graphs such as the ones derived from exponential and hyper-exponential distributions. In particular, we want to quantify the existence of strong latent associations in distance graphs. To do this, we compute the mean value of $rs$ in $TRS^{1\%}$, which we denote by $\mu$, shown in Table 8 for all distance graphs. We denote by $\pi$ the ratio of semi-metric pairs (those with $rs > 0$) from all pairs in each distance graph. This value is the percentage of semi-metric pairs, as shown in Table 6, divided by 100. We can now compare the different distance graphs according to $\mu$ and $\pi$. The first parameter quantifies the existence of strong latent associations, whereas the second quantifies the fractional or incomplete nature of a given distance graph as a representation of associative semantics. Figure 4 plots the values of each distance graph according to these two parameters.

---

[11]If we simplify this problem by considering that the distance between two vertices has equal probability of being large or small, then for a path of length 2, the probability of every segment in this path possessing a small distance is 0.25, for a path of length 3 it is 0.125, and for a path of length $n$ it is $1/n^2$. We note that for an indirect path to possess a small distance, every segment must be reasonably small.

We can clearly see in this figure that the Word Norm distance graph is the least fractional and it contains less latent associations than all the other real distance graphs. As already noted, this graph captures most semantic associations directly and those few latent associations that it entails tend not to be very strong. The same applies to the PCPAdap graph. We can say it is a good representation of the semantics of its user community. The Dissertation distance graph is very semantically complete (small percentage of semi-metric pairs), but it still contains strong latent associations amongst the few semi-metric pairs – the novel associations strongly implied by the collection of books but not made by many of the individual books (see section 5).

Table 8: μ: mean value of *rs* in *TRS*[1%].

| Graph | N | parameters | μ | Std.Dev. |
|---|---|---|---|---|
| **PCPAdap** | 423 | | 0.352776 | |
| **PCPStruct** | 423 | | 0.2007 | |
| **ISSN** | 472 | | 0.399713 | |
| **Dissert** | 86 | | 0.6080235 | |
| **ARP** | 500 | | 0.990388 | |
| **Word Norm** | 150 | | 0.319139 | |
| **Uniform** | 150 | | 0.065874 | 0.020077 |
| | 500 | | 0.051606 | 0.0051279 |
| **Exponential** | 150 | 5 | 0.166922 | 0.0166281 |
| | 500 | 5 | 0.180608 | 0.0080763 |
| | 150 | 10 | 0.319879 | 0.0327653 |
| | 500 | 10 | 0.286318 | 0.0036716 |
| | 150 | 15 | 0.3938033 | 0.0208518 |
| | 500 | 15 | 0.38278 | 0.0135159 |
| | 150 | 20 | 0.456823 | 0.0512222 |
| | 500 | 20 | 0.455522 | 0.0095474 |
| | 150 | 30 | 0.568661 | 0.0370173 |
| | 500 | 30 | 0.57756 | 0.0110469 |
| | 150 | 40 | 0.648974 | 0.0419679 |
| | 500 | 40 | 0.66449 | 0.0135768 |
| | 150 | 100 | 0.952153 | 0.0381469 |
| | 500 | 100 | 0.988852 | 0.0084951 |
| **Hyper-exponential** | 150 | (8, 39, 0.4) | 0.541686 | 0.0403997 |
| | 150 | (8, 39, 0.4) | 0.538308 | 0.014352 |
| | 150 | (8, 33, 0.5) | 0.476169 | 0.0427201 |
| | 500 | (8, 33, 0.5) | 0.469438 | 0.0038927 |
| | 150 | (9, 38, 0.1) | 0.612927 | 0.0364466 |
| | 500 | (9, 38, 0.1) | 0.61674 | 0.0099022 |

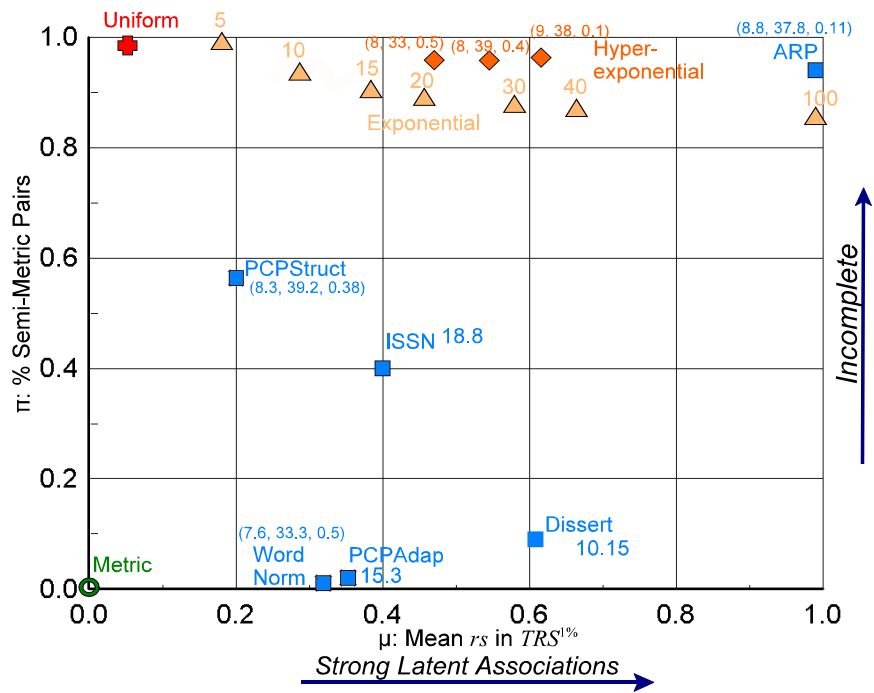

**Figure 4**: Semi-metric behavior of several distance graphs. Values for random graphs are averages; only random graphs with 500 vertices are shown.. The parameters for the exponential and hyper-exponential distributions which characterize the weights of the respective proximity graphs are also shown.

The PCPStruct and ISSN distance graphs are more fractional than the Dissertation, Word Norm and PCPAdap cases (more semi-metric pairs), but with weaker latent associations than the Dissertation. They are indeed more like random graphs than the previous ones, as they possesses plenty of semi-metric pairs, with only moderately strong latent associations. The ARP distance graph, displays high values of μ and π, showing that while it is a very incomplete representation of associative semantics, as random graphs do, it does contain many strong latent associations. In the case of ARP, this means that many strong latent associations are implied by the literature, but few articles make it explicit, suggesting plenty of room for exploiting novel associations in Science.

The three types of random graph produces are also plotted in figure 4. Uniform random graphs display a very high percentage of semi-metric pairs (fractional), but very weak latent associations. This behavior is quite distinct from the case of ARP, where we have an equally high percentage of semi-metric pairs, but very strong latent associations. However, the distribution of proximity weights in the real data sets are either exponential or hyper-exponential (see section 7.1), so we need to contrast these with more similar random graphs.

We can see in figure 4 that the exponential random graphs create a curve in the space of the two semi-metric parameters $\mu$ and $\pi$ for the parameter $\lambda$ of the exponential distribution. A non-linear regression of the values of the ratio of semi-metric pairs ($\pi$) and mean value of $rs$ in $TRS^{1\%}$ ($\mu$) yields the following model for values of $\lambda$[12]:

$$\boldsymbol{p} = a + \frac{b}{\boldsymbol{l}}$$
$$\boldsymbol{m} = c\boldsymbol{l}^2 + d\boldsymbol{l} + e \quad (7)$$

where $a$, $b$, $c$, $d$, and $e$ are real values shown in table 9 for random graphs of 150 and 500 vertices. We note that the curves for graphs of the two different dimensions are very near.

As discussed in section 7.1, the weights of the PCPAdapt, ISSN, and Dissertation proximity graphs fit an exponential random distribution with $\lambda = 15.3$; $18.8$; and $10.15$ respectively. All these three distance graphs have different sizes, N=423, 472, and 86 respectively. We compare the first two distance graphs to random graphs of 500 vertices, and the third to random graphs of 150 vertices[13]. Using formula 8 with the parameters from table 9 (those of 500 vertices for PCPAdap and ISSN, and those 150 vertices for the Dissertation data set), we obtain values of $\pi$ and $\mu$ for random counterparts of these three data sets, listed in table 10.

---

[12]The coefficient of multiple determination ($R^2$) for this model is 0.9974609914 for $\pi$ and .9881903595 for $\mu$, for graphs of 150 vertices, and 0.9881136391 for $\pi$ and 0.9961577382 for $\mu$, for graphs of 500 vertices.

Table 9: Model of semi-metric behavior of exponential random graphs

| N | a | b | c | d | e |
|---|---|---|---|---|---|
| 150 | 0.7903816582 | 0.9315307064 | -0.0000913272 | 0.01733546 | 0.1290713052 |
| 500 | 0.8550775455 | 0.7143364562 | -0.0000930467 | 0.0180254566 | 0.1147711017 |

Table 10: Comparison of real distance graphs to random counterparts

|  | $\mu$ | $\pi$ | $\mu'$random | $\pi'$random | $d$ | $d_{metric}$ | $d_{uniform}$ |
|---|---|---|---|---|---|---|---|
| **Dissert** | 0.608 | 0.09 | 0.2956 | 0.7913 | 0.7677 | 0.6146 | 1.027822 |
| **PCPAdap** | 0.3528 | 0.02 | 0.3688 | 0.9017 | 0.8818 | 0.3534 | 1.012311 |
| **ISSN** | 0.3997 | 0.4 | 0.4208 | 0.8931 | 0.4935 | 0.5655 | 0.681991 |
| **ARP** | 0.9903 | 0.94 | 0.6167 | 0.9635 | 0.3743 | 1.3654 | 0.939843 |
| **Word Norm** | 0.3191 | 0.01 | 0.4762 | 0.9794 | 0.9820 | 0.3193 | 0.986285 |
| **PCPStruc** | 0.2007 | 0.56 | 0.5383 | 0.9616 | 0.5247 | 0.5949 | 0.451777 |

As discussed in section 7.1, the PCPStruct, ARP, and Word Norm proximity graphs fit a hyper-exponential random distribution with $\lambda_1 = 8.3$, $\lambda_2 = 39.2$, $p = 0.38$; $\lambda_1 = 8.8$, $\lambda_2 = 37.8$, $p = 0.11$; and $\lambda_1 = 7.6$, $\lambda_2 = 33.3$, $p = 0.5$, respectively. Given that a hyper-exponential distribution depends on three parameters, obtaining a model curve such as the one obtained for the exponential case would require the generation of a much larger set of random graphs. Thus, we generated hyper-exponential random proximity graphs with parameters similar to those of the real data sets: $\lambda_1 = 8$, $\lambda_2 =$

---

[13]Ideally, we would compute a curve for graphs with exactly the same size of vertices as the real data sets. However, as can be seen in table 8, and also from the parameters derived for formula 8, the behavior of exponential random graphs of 150 and 500 vertices is not very distinct. Thus, our comparison of real distance graphs of size 423 and 472 to a random graph of 500 vertices, and a real graph of 86 vertices to a random graph 150 vertices is acceptable.

39, $p = 0.4$; $\lambda_1 = 9$, $\lambda_2 = 38$, $p = 0.1$; and $\lambda_1 = 8$, $\lambda_2 = 33$, $p = 0.5$, respectively. Table 10 also lists the values of $\pi$ and $\mu$ for these cases.

These values give us an indication of how far from an equivalent random graph each real distance graph is. Clearly, $\pi$ and $\mu$ give us two distinct qualities of semi-metric behavior, namely, how incomplete the associative semantics contained in a distance graph is and the strength of latent associations it contains. Above, we have already discussed these qualities for each real distance graph. But to quantify semi-metric behavior as a phenomenon, we calculate the Euclidean distance $d$ between a real distance graph and its random counterpart, as shown in table 10. As expected, the ARP, ISSN, and PCPStruct distance graphs are nearer to their random counterparts, than the other graphs. These distances can be visualized in Figure 5.

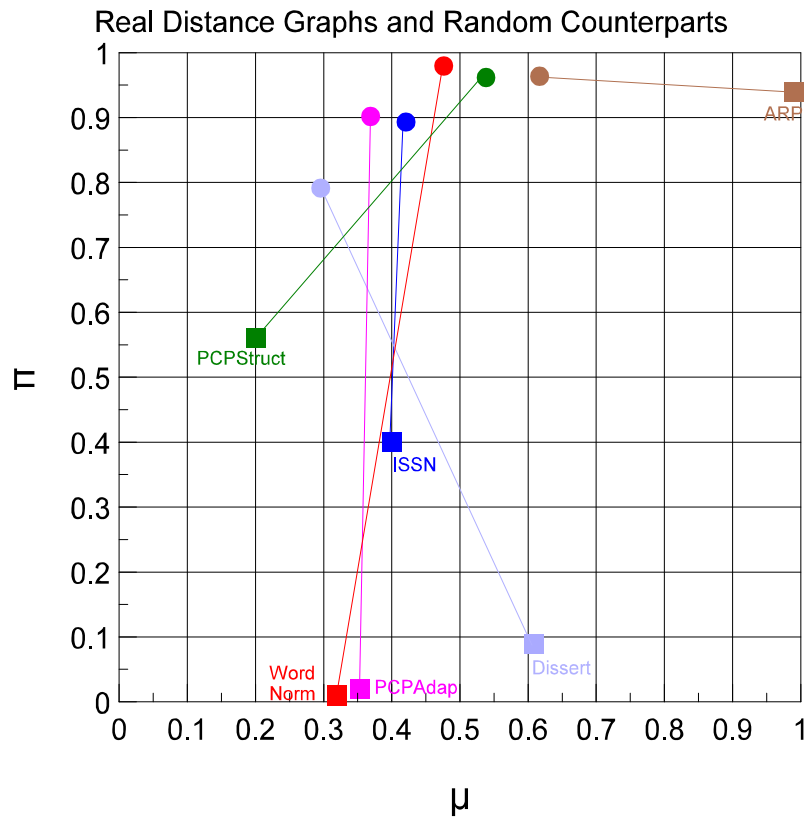

**Figure 5**: Real distance graphs (squares) and their random counterparts (circles) in $\mu$-$\pi$ space.

However, interestingly, when we calculate the Euclidean distance to mean uniform random graphs[14], $d_{uniform}$ (shown in table 10), we find that the ISSN and PCPStruct graphs are much nearer to uniform random graphs than ARP. This is a result of the high strength of latent associations in the ARP distance graph, which uniform random graphs do not observe, but which exponential and hyper-exponential random graphs observe in moderate amounts. Notice that both parameters ($\pi$ and $\mu$) would be null if we had imposed a metric distance (a transitive closure) on the original proximity graphs. Indeed, a metric distance graph would exist at the origin of the plot in figures 4 and 5. Table 10 also lists the distances of every real distance graph to a metric distance graph (the origin): $d_{metric}$.

## 8. Metric Behavior for Recommendation and Characterization of Document Networks

By utilizing a semi-metric distance function, we have gained a mechanism to classify DN according to how complete a representation of associative semantics it is, and how strong its latent associations are. Clearly, depending on the application one has in mind, the two parameters $\pi$ and $\mu$ and derivable distances, allow us to better understand a DN under study. We can detect if it reflects the expected semantics of a population, in which case it will behave more like the Word Norm and PCPAdap DN, which are nearer to metric graphs. We can also detect if a semantically complete DN still contains strong latent associations, in which case it will behave more like the Dissertation DN with high values of $\mu$ and low values of $\pi$. Conversely, a semantically incomplete DN with strong latent associations will behave more like ARP, with high values of both $\pi$ and $\mu$.

---

[14]Again, this distance is computed for graphs of 150 vertices for the Dissertation and Word Norm cases, and for graphs of 500 vertices for all other cases.

Indeed, the ability to map a semi-metric distance graph in the π/µ space, gives us a mechanism to evaluate adaptive algorithms such as the one used on the PCP Web site and the ISSN data set. Clearly Bollen's algorithm worked well on the PCP web site, as the adapted network is much closer to the Word Norm case, than the original hyperlink structure used to compute PCPStruct. But, for reasons already discussed above, it did not behave as well on the ISSN case, which was not adapted to a complete semantics nor does it entail strong latent associations.

We have compiled and presented evidence that the semi-metric analysis of distance graphs obtained from DN is a methodology useful for both recommendation of documents and chracetrization of different types of DN. Clearly, this analysis needs to be conducted for many more DN to fully develop its power, but we hope to have presented enough compelling evidence here to convince the reader that it is a methodology worth pursuing.

# 9. Integrating Evidence From Different Knowledge Contexts

## 9.1 Describing User Interest with Evidence Sets

Humans use language to communicate categories of objects in the world. But such linguistic categories are notoriously context-dependent (Lakoff 1987, Rocha 1999), which makes it harder for computer programs to grasp the real interests of users. In information retrieval we tend to use keywors to describe the content of documents, and sets of keywords to describe the present interests of a given user at a particular time (e.g. a web search).

One of the advantages of using the knowledge contexts defined in section 2 in our recommendation architecture is that the same keyterms can be differently associated in different information resources. Indeed, the distance functions of knowledge contexts allow us to regard these as connectionist memory systems (Rocha 2001a, Rocha 2001b). This way, the same set of keyterms describing the present interests (or search) of a user, is associated with different sets of other keyterms in distinct knowledge contexts. Thus, the interests of the user are also context-dependent when several information resources are at stake.

In this setting, the objective of a recommendation system that takes as input the present interest of a user, is to select and integrate the appropriate contexts, or perspectives, from the several ways the user interests are constructed in each information resource. We have developed an algorithm named *TalkMine* which implements the selective communication fabric necessary for this integration (Rocha 1999, Rocha 2001a, Rocha 2001b).

*TalkMine* uses an interval valued set structure named *evidence set* (Rocha 1994, Rocha 1999), an extension of a fuzzy set (Zadeh 1965), to model the interests of users defined as categories, or weighted sets of kewords. Evidence sets are set structures which provide interval degrees of membership, weighted by the probability constraint of the Dempster-Shafer Theory of Evidence (DST) (Shafer 1976). They are defined by two complementary dimensions: membership and belief. The first represents an interval (type-2) fuzzy degree of membership, and the second a degree of belief on that membership. Specifically, an *evidence set A* of *X*, is defined for all $x \in X$, by a membership function of the form:

$$A(x) \rightarrow (\mathcal{F}^x, m^x) \in \mathcal{B}[0, 1]$$

where $\mathcal{B}[0, 1]$ is the set of all possible bodies of evidence $(\mathcal{F}^x, m^x)$ on $\mathcal{I}$, the set of all subintervals of [0,1]. Such bodies of evidence are defined by a basic probability assignment $m^x$ on $\mathcal{I}$, for every $x$ in $X$ (see figure 6).

Each interval of membership $I_j^x$ represents the degree of importance of a particular element $x$ of $X$ (e.g. a keyterm) in category $A$ (e.g. the interests of a user) *according* to a particular *perspective* (e.g. a particular database), defined by evidential weight $m^x(I_j^x)$. Thus, the membership of each element $x$ of an evidence set $A$ is defined by distinct intervals representing different perspectives.

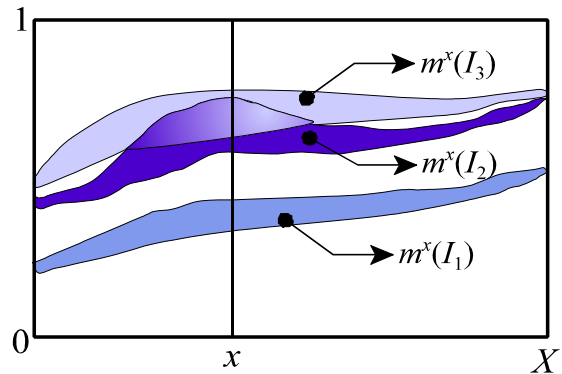

**Figure 6**: Evidence Set with 3 Perspectives

The basic set operations of complementation, intersection, and union have been defined for evidence sets, of which fuzzy approximate reasoning and traditional set operations are special cases (Rocha 1997b, Rocha 1999). Measures of uncertainty have also been defined for evidence sets. The total uncertainty of an evidence set $A$ is defined by: $U(A) = (IF(A), IN(A), IS(A))$. The three indices of uncertainty, which vary between 1 and 0, *IF* (*fuzziness*), *IN* (*nonspecificity*), and *IS* (*conflict*) were introduced in (Rocha 1997b). *IF* is based on Yager's (Yager 1979, Yager 1980) and Klir and Yuan's (Klir & Yuan 1995) measure of fuzziness. *IN* is based on the Hartley measure, and *IS* on the Shannon entropy as extended by Klir (Klir 1993) into the DST framework.

## 9.2 Inferring User Interest in Different Knowledge Contexts

Fundamental to the *TalkMine* algorithm is the integration of information from different knowledge contexts into an evidence set, representing the category of topics (described by keywords) a user is interested at a particular time. Thus, the keywords the user employs to describe her interests or in a search, need to be "decoded" into appropriate keywords for each information resource: the perspective of each knowledge context.

The *present interests* of each user can be described by a set of keywords $P^u = \{k_1, \cdots, k_p\}$. Using these keywords and the keyword distance function (2) of the several knowledge contexts involved, we want to infer the interests of the user as "seen" from the several knowledge contexts involved.

Let us assume that $r$ knowledge contexts $R_t$ are involved in addition to one from the user herself. The set of keywords contained in all the participating knowledge contexts is denoted by $\mathcal{K}$. $d_0$ is the distance function of the knowledge context of the user, while $d_1...d_r$ are the distance functions from each of the other knowledge contexts. For each knowledge context $R_t$ and each keyword $k_u$ in the user's present interests $P^u = \{k_1, \cdots, k_p\}$, a *spreading interest fuzzy set* $F_{t,u}$ is calculated using $d_t$:

$$F_{t,u}(k) = \max\left[ e^{\left(-\alpha \cdot d_t(k,k_u)^2\right)}, \epsilon \right] \quad \forall k \in R_t, t = 1...r, u = 1...p \quad (8)$$

This fuzzy set contains the keywords of $R_t$ which are closer than $\epsilon$ to $k_u$, according to an exponential function of $d_t$. $F_{t,u}$ spreads the interest of the user in $k_u$ to keywords of $R_t$ that are near according to $d_t$. The parameter $\alpha$ controls the spread of the exponential function. Because each knowledge context $R_t$ contains a different $d_t$, each $F_{t,u}$ is also a different fuzzy set for the same $k_u$, possibly even containing keywords that do not exist in other knowledge contexts. There exist a total of $n = r.p$ spreading interest fuzzy sets $F_{t,u}$ given $r$ knowledge context and $p$ keyterms in the user's present interests.

## 9.3 The Linguistic "And/OR" Combination

Since each knowledge context produces a distinct fuzzy set, we need a procedure for integrating several of these fuzzy sets into an evidence set to obtain the integrated representation of user interests we desire. We have proposed such a procedure (Rocha 2001b) based on Turksen's (Turksen 1996) combination of Fuzzy Sets into Interval Valued Fuzzy Sets (IVFS). Turksen proposed that fuzzy logic compositions could be represented by

IVFS's given by the interval obtained from a composition's Disjunctive Normal Form (DNF) and Conjucntive Normal Form (CNF): [DNF, CNF]. We note that in fuzzy logic, for certain families of conjugate pairs of conjunctions and disjunctions, DNF $\subseteq$ CNF.

Using Turksen's approach, the union and intersection of two fuzzy sets $F_1$ and $F_2$ result in the two following IVFS, respectively:

$$IV^{\cup}(x) = \left[ F_1(x) \bigcup_{DNF} F_2(x), F_1(x) \bigcup_{CNF} F_2(x) \right]$$
$$IV^{\cap}(x) = \left[ F_1(x) \bigcap_{DNF} F_2(x), F_1(x) \bigcap_{CNF} F_2(x) \right]$$
(9)

where, $A \bigcup_{CNF} B = A \cup B$, $A \bigcup_{DNF} B = (A \cap B) \cup (A \cap \overline{B}) \cup (\overline{A} \cap B)$, $A \bigcap_{CNF} B = (A \cup B) \cap (A \cup \overline{B}) \cap (\overline{A} \cup B)$, and $A \bigcap_{DNF} B = A \cap B$, for any two fuzzy sets $A$ and $B$.

Formulae (9) constitute a procedure for calculating the union and intersection IVFS from two fuzzy sets. $IV^{\cup}$ describes the linguistic expression "$F_1$ or $F_2$", while $IV^{\cap}$ describes "$F_1$ and $F_2$", – capturing both fuzziness and nonspecificity of the particular fuzzy logic operators employed, as Turksen suggested (Rocha 2001b). However, in common language, often "and" is used as an unspecified "and/or". In other words, what we mean by the statement "I am interested in x and y", is more correctly understood as an unspecified combination of "x and y" with "x or y". This is particularly relevant for recommendation systems where it is precisely this kind of statement from users that we wish to respond to.

One use of evidence sets is as representations of the integration of both $IV^{\cup}$ and $IV^{\cap}$ into a linguistic category that expresses this ambiguous "and/or". To make this combination more general, assume that we possess an evidential weight $m_1$ and $m_2$ associated with each $F_1$ and $F_2$ respectively. These are probabilistic weights ($m_1 + m_2 = 1$) which represent the strength we associate with each fuzzy set being combined. The linguistic expression at stake now becomes "I am interested in x and y, but I value x more/less

than y". To combine all this information into an evidence set we use the following procedure:

$$ES(x) = \left\{ \left\langle IV^{\cup}(x), \min(m_1, m_2) \right\rangle, \left\langle IV^{\cap}(x), \max(m_1, m_2) \right\rangle \right\} \quad (10)$$

Because $IV^{\cup}$ is the less restrictive combination, obtained by applying the maximum operator to the original fuzzy sets $F_1$ and $F_2$, its evidential weight is acquired via the minimum operator of the evidential weights associated with $F_1$ and $F_2$. The reverse is true for $IV^{\cap}$. Thus, the evidence set obtained from (10) contains $IV^{\cup}$ with the lowest evidence, and $IV^{\cap}$ with the highest. Linguistically, it describes the ambiguity of the "and/or" by giving the strongest belief weight to "and" and the weakest to "or". It expresses: "I am interested in x and y to a higher degree, but I am also interested in x or y to a lower degree".

Finally, formula (10) can be easily generalized for a combination of *n* fuzzy sets $F_i$ with probability constrained weights $m_i$:

$$ES(x) = \left\{ \left\langle IV^{\cup}_{F_i/F_j}(x), \frac{\min(m_i, m_j)}{n-1} \right\rangle, \left\langle IV^{\cap}_{F_i/F_j}(x), \frac{\max(m_i, m_j)}{n-1} \right\rangle \right\} \quad (11)$$

In *TalkMine*, this formula is used to combine the *n* spreading interest Fuzzy Sets obtained from *r* knowledge context and *p* keyterms in $P^u$ as described in section 9.2. The resulting evidence set $ES(k)$ defined on $\mathcal{K}$, represents the interests of the user inferred from spreading the initial interest set of keywords in the intervening knowledge contexts using their respective distance functions. The inferring process combines each $F_{t,u}$ with the "and/or" linguistic expression entailed by formula (11). Each $F_{t,u}$ contains the keywords related to keyword $k_u$ in the knowledge context $R_t$, that is, the perspective of $R_t$ on $k_u$. Thus, $ES(k)$ contains the "and/or" combination of all the perspectives on each keyword $k_u \in \{k_1, \cdots, k_p\}$ from each knowledge context $R_t$.

As an example, without loss of generality, consider that the initial interests of an user contain one single keyword $k_1$, and that the user is querying two distinct information resources $R_1$ and $R_2$. Two spreading interest fuzzy sets, $F_1$ and $F_2$, are generated using $d_1$ and $d_2$ respectively, with probabilistic weights $m_1=v_1$ and $m_2=v_2$, say, with $m_1 > m_2$ to indicate that the user trusts $R_1$ more than $R_2$. $ES(k)$ is easily obtained straight from formula (10). This evidence set contains the keywords related to $k_1$ in $R_1$ "and/or" the keywords related to $k_1$ in $R_2$, taking into account the probabilistic weights attributed to $R_1$ and $R_2$. $F_1$ is the perspective of $R_1$ on $k_1$ and $F_2$ the perspective of $R_2$ on $k_1$.

# 10. Distance Functions in Recommendation Systems

The evidence set combination defined in Section 9.3 with formulas (10) and (11) is a first cut at detecting the interests of a user in a set of information resources. Our *TalkMine* recommendation algorithm computes a more tuned interest set of keywords, using an interactive conversation process between the user and the information resources being queried. Such conversation is an uncertainty reducing process based on the IR system of Nakamura and Iwai (Nakamura & Iwai 1982), which we extended to Evidence Sets (Rocha 1999, Rocha 2001b).

*TalkMine* can be understood as an algorithm for obtaining a representation of user interests in several information resources (including other users). It works by combining into an evidence set, the present user interests with all the perspectives derived from each information resource. The resulting Evidence Set is further fine-tuned by an automated conversation process with the user's agent/browser (Rocha 2001b) . The combination of perspectives utilizes the evidence set combination defined in section 9, which in turn employs the semi-metric distance functions described in this article. The importance of such semi-metric distance functions on their own, is also described in this article. They allow us to both characterize Document Networks for interests and trends (useful for recommendation), as well as offer an avenue to combine user interests in distinct information resources.

In this article we have detailed empirical evidence of the utility of semi-metric distance functions for recommendation processes. In particular, we emphasize that forcing distance functions in recommendation systems to be metric, leads to the loss of important information entailed in DN. Namely, the capacity to identify strong latent associations, trends, and to characterize and compare different DN. We have also offered a mechanism to integrate associations (defined by distances) of items from different DN into a single representation (an evidence set) useful for recommendation processes.